\documentclass[preprint2]{aastex}

\newcommand{\rosat}{{\it ROSAT}}
\newcommand{\asca}{{\it ASCA}}

\newcommand{\cxo}{{\it Chandra}}

\newcommand{\nh}{\mbox {$N_{\rm H}$}}

\newcommand{\hi}{H\,{\sc i}}

\newcommand{\about}{$\sim$\kern.03em}
\newcommand{\ergs}{ erg s$^{-1}$}

\newcommand{\g}{G18.95--1.1}

\slugcomment{Draft \today}

\shorttitle{Indications for a pulsar and its wind nebula in \g}
\shortauthors{T\"ullmann et al.}

\begin{document}


\title{Searching for the pulsar in \g: Discovery of an X-ray point source and associated synchrotron nebula with Chandra}


\author{R. T\"ullmann\altaffilmark{1},
P. P. Plucinsky\altaffilmark{1},
T. J. Gaetz\altaffilmark{1},
P. Slane\altaffilmark{1},
J. P. Hughes\altaffilmark{2}, 
I. Harrus\altaffilmark{3,*}, and
T. G. Pannuti\altaffilmark{4}
}

\altaffiltext{1}{Harvard-Smithsonian Center for Astrophysics,
60 Garden Street, Cambridge, MA 02138; rtuellmann@cfa.harvard.edu}
\altaffiltext{2}{Department of Physics and Astronomy, Rutgers University, 136 Frelinghuysen Road, Piscataway, NJ 08854}
\altaffiltext{3}{NASA/GSFC}
\altaffiltext{*}{Currently at NASA HQ, Washington DC}
\altaffiltext{4}{Space Science Center, 235 Martindale Drive, Morehead State University, Morehead, KY 40351}

\begin{abstract}
Using the \cxo\ X-ray Observatory, we have pinpointed the location of a faint X-ray point source (CXOU\,J182913.1-125113) and an associated diffuse nebula in the composite supernova remnant G18.95-1.1. These objects appear to be the long-sought pulsar and its wind nebula. The X-ray spectrum of the point source is best described by an absorbed powerlaw model with $\Gamma=1.6$ and an $N_H$ of $\sim$$1\times10^{22}$\,cm$^{-2}$. This model predicts a relatively low unabsorbed X-ray luminosity of about $L_X (0.5-8.0{\rm keV})\simeq4.1\times10^{31}$\,$D_2^2$\,erg s$^{-1}$, where $D_2$ is the distance in units of 2kpc. The best-fitted model of the diffuse nebula is a combination of thermal ($kT=0.48$keV) and non-thermal ($1.4\le\Gamma\le1.9$) emission. The unabsorbed X-ray luminosity of $L_X\simeq5.4\times10^{33}$\,$D_2^2$\,erg s$^{-1}$ in the 0.5\,--\,8keV energy band seems to be largely dominated by the thermal component from the SNR, providing 87\% of $L_X$ in this band. 
No radio or X-ray pulsations have been reported for CXOU\,J182913.1-125113. If we assume an age of $\sim$5300\,yr for G18.95-1.1 and use the X-ray luminosity for the pulsar and the wind nebula together with the relationship between spin-down luminosity (via magnetic dipole radiation) and period, we estimate the pulsar's period to be $P\simeq0.4$\,s. Compared to other rotation-powered pulsars, a magnetic field of $2.2\times10^{13}$\,G is implied by its location in the $P$--$\dot P$ diagram, a value which is close to that of the quantum critical field.
\end{abstract}



%% In the first two sections, notice the use of athe natbib \citep

\section{Introduction}
Supernova remnants (SNRs) with a center-filled or composite morphology often host young pulsars whose strong relativistic winds can lead to the formation of pulsar wind nebulae (PWNe). In recent years the high spatial resolution of \cxo\ has made it possible to detect the PWNe, to pinpoint the locations of the compact objects, and to constrain their nature. A new, yet unproven, case in which a putative pulsar is suspected to power a PWN is \g. For this composite SNR, which is considered to be located in the Sagittarius Arm at a distance of $\sim$\,2\,kpc \citep{fuerst89}, an X-ray luminosity of $6.1\times10^{34}$\,erg s$^{-1}$ (0.1\,--\,10\,keV) has been reported by \citet{harrus04}. \g\ was discovered as a shell-like non-thermal extended radio source \citep{reich84}. Further observations carried out at GHz frequencies by \citet{fuerst85}, \citet{ode86}, and \citet{fuerst97} detected an elongated synchrotron nebula in the center of \g\ and a region of enhanced radio emission at its tip, much like the central PWN in W44 \citep[e.g.,][]{petre02}. Another region of enhanced radio emission is located along the western shell of \g.  The age of the remnant and its radio diameter are considered to be (4.4\,--\,$6.1)\times10^3$\,yr \citep{harrus04} and $\sim$33$\arcmin$ \citep[e.g.,][]{ode86,patna88,fuerst89} which is equivalent to $\sim$18$D_2$\,pc, where $D_2$ is the distance in units of 2kpc.
\begin{figure*}[!pht]
\centering
\includegraphics[width=16.25cm,height=8cm,clip]{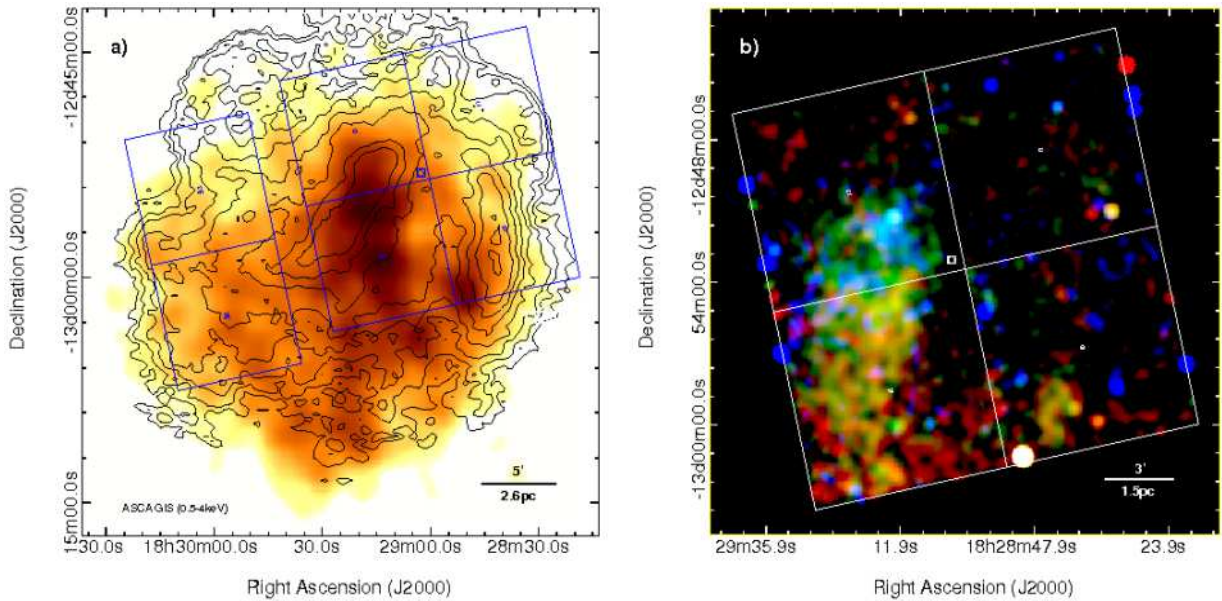}
\caption{\label{f1} a): X-ray image of \g\ obtained with the \asca\ GIS ($FWHM$=45\arcsec) in the 0.5\,--\,4keV energy band and overlaid with 10.55GHz radio continuum contours and the {\tt ACIS-I} FOV. b:) Exposure-corrected multi-color X-ray image of the inner part of \g\ (red\,=\,0.35\,--\,1.1\,keV, green\,=\,1.1\,--2.6\,keV, blue\,=\,2.6\,--\,8.0\,keV). The image has been binned by a factor of 8 and smoothed with a Gaussian kernel of $FWHM$\,=\,6\,pix.}
\end{figure*}

In a previous X-ray study carried out with \rosat\ and \asca, \citet{harrus04} mapped the diffuse X-ray emission of \g. A reproduction of the \asca\ image in the 0.5\,--\,4keV energy band is shown in Fig.~\ref{f1}a together with radio contours at 10.55GHz \citep{fuerst97}. The radio emission consists of a central elongated radio-bright nebula \citep[$L_{10^7-10^{11}{\rm Hz}}=5\times10^{33}$\ergs,][]{fuerst89} running from the NW to the SE and another region of enhanced radio emission located at the western shell of \g. \citet{harrus04} also detected a localized region of hard X-ray emission in the 4\,--\,10keV energy band whose peak emission is centered at RA=18$^h$29$^m$03\fs3, Dec=$-12^{\circ}$52\arcmin58\farcs9 (see also their Fig.~2 for its location). This hard X-ray emission is slightly offset from the region of enhanced radio synchrotron emission, similar to G326.3--1.8 \citep{plu02}. Harrus et al. found that the X-ray emission from \g\ is predominantly thermal, heavily absorbed with a column density of about $1\times10^{22}$\,cm$^{-2}$, and can be best described by a non-equilibrium ionization (NEI) model with a temperature of $\sim$0.9\,keV and an ionization timescale of $1.3\times10^{10}$\,cm$^{-3}$\,s$^{-1}$. In a reanalysis of the \asca\ data of the localized hard X-ray emission, we confirmed the existence of a non-thermal component, which seems to dominate the spectrum above 2\,keV. Although \asca's spatial resolution is insufficient to detect a point source within this region, it is possible that the hard X-ray emission indicates the location of an undetected pulsar.

We have learned from \cxo\ observations that many young pulsars, like those in 3C 58 and G21.5-0.9 as well as currently undetected pulsars, e.g., the one in G327.1-1.1, are embedded in extended non-thermal emission from the PWNe. In the case of \g\ the extended emission is most likely also contaminated by diffuse thermal emission from the SNR, but any non-thermal emission from a possible point source remained unresolved by \asca. With \cxo\ we now have the opportunity to detect a point source in this extended emission. Therefore, we conducted a follow-up \cxo\ observation of G18.95-1.1 in order to search for an X-ray point source embedded in the extended emission, to determine its nature, and to characterize the spatial and spectral distribution of the hard emission to test if it is consistent with an X-ray synchrotron nebula.

\section{Follow-up \cxo\ Observations}
A 45\,ks \cxo\ observation was executed on August 3rd, 2009 (ObsID 10098) using {\tt ACIS-I} in {\tt vfaint} mode with the aimpoint of the {\tt ACIS-I} array centered onto the region in which the 4\,--\,10keV emission from \asca\ overlapped the radio synchrotron emission (see Fig.~\ref{f1}a). Data reduction was carried out with {\tt CIAO4.2} and {\tt CALDB} version 4.2.2, following standard procedures\footnote{http://cxc.harvard.edu/ciao/guides/acis\_data.html}, except for the charge transfer inefficiency (CTI) correction. 
As during the last 11ks of the observation the focal plane temperature was warmer than -118.7C (increasing from -119.0C to -118.4C), we applied a time-dependent CTI correction following \citet{posson10}. 
The effects of this correction are marginal in the 0.35\,--\,8.0keV energy band as it adjusted the gain on average by $\sim$0.5\% to account for the increased CTI at higher temperatures.
%\begin{figure}[t]
%\centering
%\includegraphics[width=5cm,height=8cm,clip,angle=-90]{f3.ps}
%\caption{\label{f3} Radial profiles of the simulated point source (solid line) and the point source candidate, providing evidence that this X-ray source is indeed a point source.  
%}
%\end{figure}

\begin{figure}[t]
\centering
\includegraphics[width=7.5cm,height=7.4cm,clip]{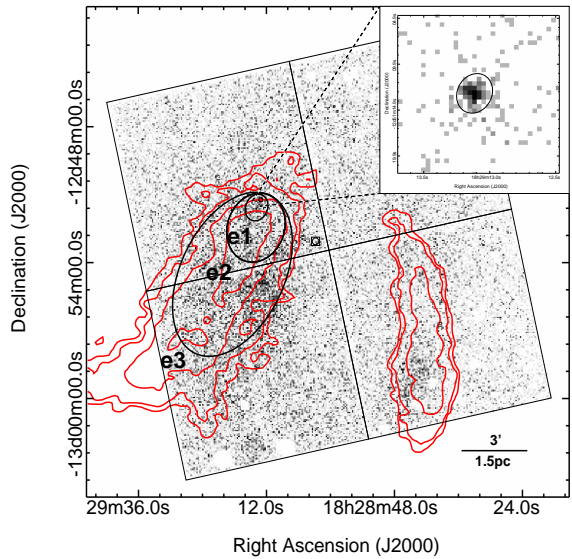}
\caption{\label{f2} 10.55GHz radio continuum contours \citep{fuerst97} overlaid onto a \cxo\ counts image (bin=8, 0.35\,--\,2.6keV, point source-subtracted). The inlay shows the unbinned counts image of the point source in the full band and a {\tt SAOTrace} simulation of the PSF (90\% encircled energy). Both images are displayed on a square root scale and plotted between 0 and 12 counts\,pix$^{-1}$. Spectral extraction regions are shown as black ellipses.}
\end{figure}
\section{Data Analysis}
In Fig.~\ref{f1}b we show the high-resolution \cxo\ X-ray composite image of the inner part of \g. It reveals a region of hard X-ray emission at RA=18$^h$29$^m$13\fs1, Dec=$-12^{\circ}$51\arcmin13\farcs4 ($l$=$+18^{\circ}$58\arcmin58\farcs4, $b$=$-00^{\circ}$59\arcmin59\farcs1), which is shown in greater detail in Fig.~\ref{f2}. This hard emission seems to originate from a point-like source and its immediate vicinity (its extent corresponds to region e1 highlighted in Fig.~\ref{f2}). The point-like source is located at the northern tip of a trail of softer diffuse X-ray emission extending towards the SE and pointing back towards the center of the remnant. The extended emission is well aligned with the radio synchrotron nebula, but does not appear to cover the full extent of the latter (see red contours in Fig.~\ref{f2}). 

Comparing the radial profile of this source with the one from an {\tt SAOTrace} simulation, confirmed that this source is indeed a point source. This object, CXOU\,J182913.1-125113 hereafter, is the only X-ray point source detected within a circle of 1\farcm4 radius centered on the position of the source. CXOU\,J182913.1-125113 has no known radio, 2MASS, optical, or $\gamma$-ray counterpart. 

The brightest point source in the field (the bright white object near the southern edge of the FOV in Fig.~\ref{f1}b) was classified by \citet{harrus04} as a late-type star and appears not to be associated with \g. There are other sources in the field which could be pulsar candidates, notably the pair of sources near RA\,=\,18$^h$28$^m$36\fs0, Dec\,=\,$-12^{\circ}$51\arcmin00\farcs8 ($l$\,=\,$+18^{\circ}$54\arcmin57\farcs4, $b$\,=\,$-00^{\circ}$51\arcmin52\farcs0). However, these sources have soft X-ray spectra as indicated by their colors in Fig.~\ref{f1}b, are $\sim 8.0\arcmin$ away from the diffuse nebula, and can be associcated with foreground stars in the USNO-b1.0 catalog (Monet et al. 2003). 
The other point-like objects in the field which appear on or outside the edge of the FOV are most likely artifacts of the smoothing in regions with low and/or variable exposure. In addition, faint enhancements in the diffuse emission or in the background might appear to be point-like after smoothing.

Although the searches for pulsations in the radio by \citet{fuerst89} proved unsuccessful (their closest search position is $\sim$$3\farcm4$ away from CXOU\,J182913.1-125113), the coincidence of the X-ray and radio synchrotron emission as well as the cometary-shaped morphology of the diffuse X-ray emission, with a hard X-ray point source at the tip, seems to point at a pulsar and its wind nebula.

To test if the point source and the extended emission are consistent with this interpretation, we extracted spectra for the putative pulsar and from three extended regions to the south of this source, using the CIAO tool {\tt specextract}.
For the putative pulsar we extracted all counts within a radius of $5\farcs2$, while the three elliptical extraction regions shown in Fig.~\ref{f2} were used for the extended emission from which only the point source was excluded. Two additional spectra were extracted which cover different fractions of the extended X-ray emission. They contain all counts in the e2\,$-$\,e1 and e3\,$-$\,e2 regions. This approach was chosen because we wanted to see how the spectra and the X-ray luminosity change as a function of increasing distance from the pulsar candidate.
Background for the pulsar candidate was taken from an annular region centered on the position of this source and from regions close to the ACIS chip edges for the extended X-ray emission. 
All spectra were grouped with {\tt dmgroup} until at least a signal-to-noise ratio of 3 (5) was achieved for the pulsar candidate (extended sources). 

\begin{figure}[t]
\centering
\includegraphics[width=7.5cm,height=12cm,clip]{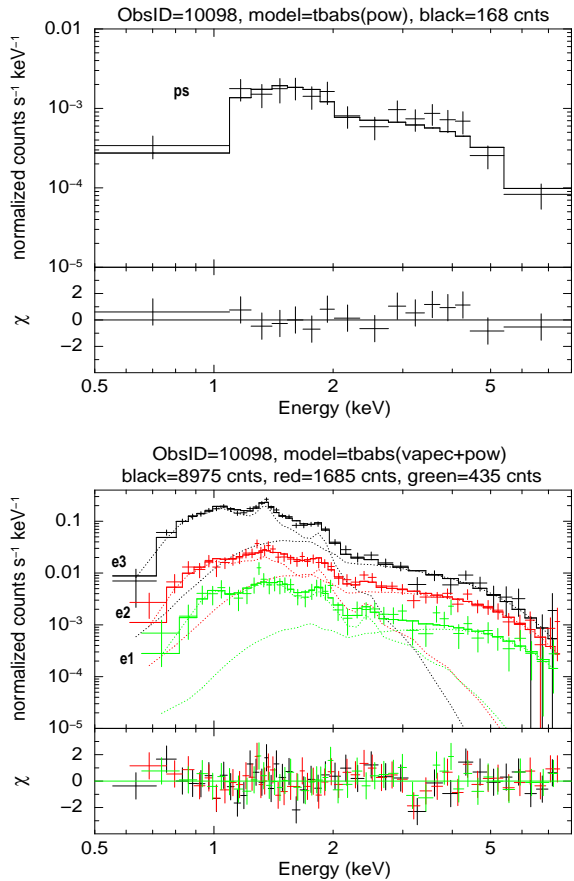}
\caption{\label{f4} Top: Best-fitted powerlaw model for the point source (ps). % and the extended region north of the pulsar (b1). 
Bottom: Spectra extracted from regions e1 to e3 can be best fitted by a combination of a $vapec$ model and a powerlaw.}
\end{figure}

\begin{deluxetable}{ccccccccccc}
\tabletypesize{\footnotesize}
\tablecolumns{10}
\tablewidth{0pt} 
\tablecaption{Best-fit results for the purely diffuse emission}
\tablehead{
\colhead{region}                                     &
\colhead{$\chi^{2}_{\rm red}/\nu$}                   & 
\colhead{$P$ value}                                  & 
\colhead{$K_1$}                                      & 
\colhead{$K_2$}                                      & 
\colhead{$N_{\rm H}$}                                & 
\colhead{$kT_1$}                                     & 
\colhead{$\Gamma$}                                   & 
\colhead{$L_{\rm 0.5-8}$}                            & 
\colhead{$L_{\rm 2.0-8.0}$}                          \\
\colhead{(1)} & 
\colhead{(2)} & 
\colhead{(3)} & 
\colhead{(4)} & 
\colhead{(5)} & 
\colhead{(6)} & 
\colhead{(7)} & 
\colhead{(8)} & 
\colhead{(9)} & 
\colhead{(10)} 
}
\startdata                                                                                           
e1 & 0.81/44 & 9.21E-01 & 4.74E-04 & 2.13E-05 & 1.97$^{+0.52}_{-0.56}$ & 0.48 (frozen)          & 1.4$^{+0.2}_{-0.6}$ & 3.72$^{+0.54}_{-0.24}$ & 0.60$^{+0.36}_{-0.12}$ \\
e2 & 0.71/47 & 9.11E-01 & 4.53E-04 & 1.62E-04 & 1.56$^{+0.36}_{-0.64}$ & 0.48 (frozen)          & 1.9$^{+0.3}_{-0.4}$ & 9.36$^{+2.32}_{-1.60}$ & 1.76$^{+1.56}_{-1.12}$ \\
e3 & 0.91/35 & 6.22E-01 & 7.31E-03 & 2.45E-04 & 1.33$^{+0.12}_{-0.13}$ & 0.48$^{+0.06}_{-0.12}$ & 1.8$^{+0.6}_{-1.0}$ & 54.4$^{+8.7}_{-4.9}$   & 4.40$^{+4.32}_{-0.92}$ \\[0.1cm]
\hline \\[-0.2cm]
e2\,--\,e1 & 0.82/47 & 8.04E-01 & 1.20E-03 & 7.23E-05 & 1.54$^{+0.25}_{-0.32}$ & 0.48 (frozen)          & 1.7$^{+0.4}_{-0.4}$ & 9.68$^{+1.88}_{-0.84}$ & 1.28$^{+1.24}_{-0.20}$ & \\
e3\,--\,e2 & 1.59/22 & 3.90E-02 & 5.35E-03 & 1.31E-04 & 1.30$^{+0.12}_{-0.14}$ & 0.46$^{+0.06}_{-0.15}$ & 1.9$^{+1.4}_{-1.1}$ & 37.5$^{+5.2}_{-3.4}$ & 2.12$^{+2.16}_{-0.40}$ & \\
\enddata
\vspace{-0.4cm}
\tablenotetext{\mbox{}}{(1): Extraction regions as shown in Fig.~\ref{f2}. (2): Reduced $\chi^2$ of the best fit/degrees of freedom. (3): Assuming that the model is a good fit to the data, the $P$ value is the probability of getting a $\chi^2$ which is greater than the observed one. (4) and (5): $K_{1,2}$ are the normalization constants of the additive model components in units of cm$^{-5}$ for $K_1$ ($vapec$) and photons keV$^{-1}$\,cm$^{-2}$\,s$^{-1}$ at 1\,keV for $K_2$ ($pow$). (6): HI column density in units of $10^{22}$\,cm$^{-2}$. (7): Temperature of the thermal plasma component. (8): Photon index. (9) and (10): Unabsorbed X-ray luminosities in units of $10^{32}$\,erg\,s$^{-1}$. }
\label{t1}
\end{deluxetable}

\subsection{The X-ray Point Source}
The spectrum labeled 'ps' in Fig.~\ref{f4} shows the X-ray emission from the point source. It contains 168 net counts and was fitted in {\tt XSPEC} (v12.6.0) with an absorbed powerlaw model and alternatively with an absorbed thermal plasma model. To simulate the absorption the $tbabs$ model \citep{wilms00} was used while the thermal emission was modeled with the $apec$ model \citep{smith01}. The $N_H$ was allowed to vary, but was initially set to the Galactic value of $N_H=(1.3\pm0.4)\times 10^{22}$\,cm$^{-2}$ \citep{dl90}.

The best-fitted model ($\chi^2/\nu=12.87/35=0.37$) is the absorbed powerlaw with a photon index of $\Gamma=1.6^{+0.4}_{-0.4}$ and an $N_H=9.5^{+6.5}_{-4.7}\times 10^{21}$\,cm$^{-2}$ (all uncertainties are given for a 90\% confidence level). The predicted $N_H$ is in good agreement with previous measurements from \citet{fuerst97} and \citet{harrus04}.
 Although the thermal plasma model provides an equally good fit, the predicted $kT$ of 16keV is unreasonably high and therefore makes a thermal nature of the source very unlikely. 
The unabsorbed X-ray luminosities predicted by the powerlaw model in the 0.5\,--\,8.0keV and 2.0\,--\,8.0keV energy bands amount to $4.08^{+3.32}_{-1.00}\times 10^{31}$\,$D_2^2$\ergs\ and $2.12^{+2.68}_{-1.40}\times 10^{31}$\,$D_2^2$\ergs, respectively. The powerlaw index of 1.6 is consistent with the pulsar hypothesis and the derived luminosity is towards the low end of the distribution of pulsar luminosities. This may not be surprising since the brighter objects have already been discovered.

Under the simplistic assumption that the stellar remnant traverses the distance from the geometric center of the SNR (to be assumed at RA=18$^h$29$^m$23\fs9, Dec=$-12^{\circ}$59\arcmin05\farcs1 or equivalently $l$=$+18^{\circ}$53\arcmin13\farcs6, $b$=$-01^{\circ}$05\arcmin57\farcs7) to its current location (4.3pc $D_2^2$) at a constant speed within 4400 to 6100 years, transverse velocities between 700\,km\,s$^{-1}$ and 960\,km\,s$^{-1}$ are estimated. Although these values are significantly higher than the typical ones of about 500\,km\,s$^{-1}$ \citep{gaensler06}, they are not unusual for pulsars with periods significantly larger than several milliseconds \citep{toscano99}. A pulsar would be consistent with the derived spectral index from the powerlaw fit and the displacement between the radio synchrotron nebula and the X-ray point source.

\subsection{The PWN candidate}
The spectra of the purely diffuse X-ray emission extracted from regions e1, e2, and e3 are shown in the lower panel of Fig.~\ref{f4} together with the best-fitted model ($tbabs(vapec+pow)$). Fitting the spectra with a single VNEI model \citep{bork01} resulted in $\chi^2_{\rm red}$ values $>$2, while a VNEI model with an additional component (e.g., $tbabs(vnei+pow)$) produced ionization timescales of $(2.3\pm1.1)\times 10^{13}$\,s\,cm$^{-3}$ which indicate that the plasma is in collisional ionization equilibrium.

The spectrum labeled e3 covers the bulk of the extended X-ray emission near the center of \g. It has 8975 net counts and provides evidence of \ion{Mg}{11} emission around 1.35keV. The model predicts a thermal plasma temperature of 0.48keV and a photon index of $\Gamma$\,=\,1.8. The best-fit results are shown in Table~\ref{t1} and are given for a 90\% confidence level. In case of e3, we freed the magnesium abundance to remove significant residuals around this line, yielding a relative Mg abundance of 1.4$\pm0.3$. This ratio is consistent with the Galactic value and agrees well with that reported by \citet{harrus04}. The X-ray luminosities for e3 provide evidence that the gas is predominantly thermal, as 87\% of $L_X$ is emitted by the thermal plasma component in the 0.5\,--\,8keV energy band. The thermal emission could originate from the SNR's shell seen along the line-of-sight or from the PWN if it has been disturbed by the reverse shock so that thermal emission and non-thermal emission are intertwined.

For e3, as well as for e1 and e2, the fitted \hi\ column density is consistent with the average Galactic value of $(1.3\pm0.4)\times10^{22}$\,cm$^{-2}$. A column that high seems to imply a distance much larger than 2\,kpc. However, as \citet{fuerst89} and \citet{harrus04} argue, the alternative distance of 15\,kpc suggested by \citet{fuerst89} appears unlikely, as the derived radio luminosity at this distance would be twice that of the Crab nebula, and the swept up mass of about 3000$M_{\odot}$ and the explosion energy of $2\times10^{52}$\,erg are unreasonably high. 
Even if we determine a distance based on the best-fit \nh, assuming $\nh=1.3\times10^{-22}$\,cm$^{-2}$ and an \hi\ volume density of $0.4$\,cm$^{-3}$ \citep{naka03}, yielding a distance of about 11\,kpc, the explosion energy of $9\times10^{51}$\,erg as well as the swept up mass of $\approx$$1400$\,M$_{\odot}$ are still unreasonably high. We therefore adopt a distance of 2\,kpc for \g\ and quote our results as a function of distance.

The \ion{Mg}{11} line emission seen in e3 remains undetected in the spectra for regions e2 and e1 (1685 and 435 net counts, respectively), but the bumps at about 1.87keV indicate the presence of \ion{Si}{13} emission and therefore a partly thermal origin of the gas. The photon index does not steepen significantly from e1 to e2 and the emission remains soft. For e1 73\% and for e2 51\% of the X-ray luminosity is produced by the thermal plasma component.

Fitting the spectra e1, e2, and e3 with only an absorbed powerlaw generally produces significantly worse fits ($0.78\le\chi^2_{red}\le4.02$) with photon indices of $\Gamma$\,=\,1.9 (e1), 2.2, (e2), and 3.9 (e3), respectively. Moreover, the potential emission features make a pure non-thermal origin of the X-ray emission appear very unlikely.

Because the pulsar and the PWN candidate are embedded within \g, the non-thermal emission from these objects is expected to be highly contaminated by the thermal emission from the SNR. We therefore adopt the temperature obtained from spectrum e3 to be representative for the inner part of \g\ and freeze the $kT$ for e1, e2, and $e2-e1$, respectively (see Table~\ref{t1}). This approach seemed to be justified, because spectra extracted from diffuse emission north of the pulsar candidate also suggested the $kT$ to range from about 0.40 to 0.55\,keV with relatively large uncertainties of 0.15\,keV and \nh\ values which are consistent with those listed in Table~\ref{t1}.

For the extended X-ray emission to be consistent with a PWN, several criteria have to be met \citep[see e.g.][]{safi04,gaensler06}. First, the X-ray surface brightness should decrease with increasing distance from the point source. Second, the radio spectral index $\alpha$ should be relatively flat ($0\le\alpha\le0.3$, $S_{\nu}\sim \nu^{-\alpha}$). Third, the photon index $\Gamma$ should be steeper at X-ray energies than in the radio.  As a consequence, the extent of the X-ray emission can be smaller than the extent of the radio emission, as the synchrotron lifetime of the electrons in the X-ray regime can be shorter than those at radio wavelengths. 

To test these criteria and to model how much the spectra change with increasing distance from the pulsar, we discuss spectra extracted from regions e1, $e2-e1$, and $e3-e2$ and focus on the hard, non-thermal X-ray emission emitted above 2keV, which we solely attribute to the putative pulsar and the PWN.
As can be seen from Table~\ref{t1}, the fraction of the X-ray luminosity above 2keV decreases from 16\% (e1), to 13\% ($e2-e1$), down to 6\% ($e3-e2$). If we take the hard X-ray luminosities and divide by the area of the extraction regions, we get surface brightnesses which decrease from $(1.77^{+1.06}_{-0.35})\times10^{28}$\,$D_2^2$\,erg\,s$^{-1}$\,arcsec$^{-2}$ for e1, to $(6.60^{+6.40}_{-0.35})\times10^{27}$\,$D_2^2$\,erg\,s$^{-1}$\,arcsec$^{-2}$ for $e2-e1$, down to $(2.64^{+1.54}_{-0.05})\times10^{27}$\,$D_2^2$\,erg\,s$^{-1}$ arcsec$^{-2}$ for $e3-e2$. Hence, the first criterion is fulfilled. 

For the inner region of \g, flat radio spectral indices of $\alpha=0.25\pm0.12$ \citep{ode86} and $\alpha=0.26\pm0.05$ \citep{fuerst89} have been reported. As the radio emission is also polarized in this region, \citet{fuerst85} report an integrated polarization percentage of about 2.5\% at 4.75GHz for the whole remnant, the second criterion is met, too. The average photon index derived from the X-ray spectral fits is $\Gamma=1.7^{+0.5}_{-0.4}$, significantly steeper than the radio index, thereby satisfying the third criterion. There is no significant steepening towards the edge of the nebula (see column 8, Table~\ref{t1}). We also fitted the spectra with the \nh\ fixed to the value of e3, but the uncertainties were still too large to say with any confidence that there is a clear trend in photon index. Furthermore, the major half-axis diameter of the X-ray emission in region e3 is about $7\farcm7$ ($\approx$\,4pc\,$D_2$), which is much less than the extent of the radio synchrotron contours at 10.55GHz in Fig.~\ref{f2} indicate. We conclude, that the extended hard X-ray emission originates from a PWN with a photon index between $1.4\le\Gamma\le1.9$ and an unabsorbed luminosity of $L_{X}=4.4\times10^{32}$\,$D_2^2$\ergs\ in the 2\,--\,8keV energy band.

\section{Discussion}\label{sec-discussion}
We can now begin to estimate some of the fundamental properties of the putative pulsar in G18.95-1.1. Using the relationship between the spin-down energy loss rate ($\dot E$) and the non-thermal X-ray luminosity of the pulsar and the PWN \citep{possenti02} of $(4.6^{+4.3}_{-0.9})\times10^{32}$\,$D_2^2$\ergs, we estimate $\dot E$ to be $(6.7^{+6.4}_{-2.0})\times10^{35}$\ergs. Following \citet{seward88}, adopting an average age of \g\ of $(5300\pm900)$\,yr \citep[see][]{harrus04}, a canonical moment of inertia of the neutron star of $I=10^{45}$\,g\,cm$^2$, and assuming further a constant braking index of 3, that spin down happens via magnetic dipole radiation, and that the current value of the pulsar's spin period is much larger than its initial one, the current spin period ($P$), the period derivative ($\dot P$), and the surface magnetic field of the pulsar is estimated to be $P=0.4^{+0.4}_{-0.3}$\,s, $\dot P=(1.2^{+0.9}_{-0.8})\times 10^{-12}$\,s\,s$^{-1}$, and $B_0=(2.2^{+1.9}_{-1.6})\times 10^{13}$\,G, respectively. These parameters can only serve as rough estimates as they are derived from the relation between $L_X$ and $\dot E$ which can vary by about one order of magnitude \citep[see e.g.,][]{possenti02}. It should also be kept in mind that $B_0$ gives the field strength at the magnetic equator and that the field strength at the magnetic poles can be a factor of 2 higher.

If the inferred values for the period, period derivative, and magnetic field are correct, this pulsar candidate occupies a highly interesting and sparsely populated place in the $P-\dot P$ diagram. It would be located close to the line of the quantum critical field of $B_{c} = 4.4\times10^{13}$\,G \citep[see e.g.,][]{kondra09}, which divides normal radio pulsars and the more exotic radio-quiet objects, such as the anomalous X-ray pulsars (AXP) and soft gamma-ray repeaters (SGRs). %This field currently receives much attention due to the discovery of several radio-quiet pulsars with magnetic fields comparable to or higher than the quantum critical field. Moreover, until recently, AXPs were not known to have PWNe, but very recent observations indicate that at least one AXP possesses associated extended X-ray emission \citep{vinka09}, which makes the case of G18.95-1.1 even more interesting.

\section{Summary}\label{sec-summary}
We report the detection of a faint X-ray point source, CXOU\,J182913.1-125113, in the inner part of the composite SNR \g\ and a trail of diffuse emission pointing towards the center of the SNR. These sources are considered to be the pulsar and its wind nebula. The best fit to the spectrum of the pulsar candidate is an absorbed powerlaw model ($\Gamma$\,=\,1.6, \nh\,=\,$1\times10^{22}$\,cm$^{-2}$) with an unabsorbed luminosity in the 0.5\,--\,8keV energy band of $L_X$\,$\simeq$\,$4.1\times10^{31}$\,$D_2^2$\ergs. The X-ray luminosity of the PWN ($1.4\le\Gamma\le1.9$, \nh\,=\,$1.6\times10^{22}$\,cm$^{-2}$) amounts to $L_X$\,$\simeq$\,$5.4\times10^{33}$\,$D_2^2$\ergs\ in the 0.5\,--\,8keV energy band. It appears as if the spectrum extracted from the region which contains the PWN shows thermal emission from the SNR along the line-of-sight or the PWN has been disturbed by the reverse shock so that thermal and non-thermal emission is mixed.

Although pulsations have not been reported for the pulsar, we estimate the spin period and the period derivative to be $P$\,=\,$0.4^{+0.4}_{-0.3}$\,s and $\dot P$\,=\,$(1.2^{+0.9}_{-0.8})\times 10^{-12}$\,s\,s$^{-1}$, assuming the dipole approximation. Compared to other rotation-powered pulsars, a high magnetic field of $B_0=(2.2^{+1.9}_{-1.6})\times 10^{13}$\,G is implied by its location in the $P$\,--\,$\dot P$ diagram. This value is close to that of the quantum critical field and makes this source a highly interesting object for follow-up observations.

\acknowledgments
Support for this work was provided by NASA through \cxo\ Award Number GO9-0081X. %We thank the anonymous referee for useful comments.% issued by the \cxo\ X-ray Observatory Center, which is operated by the Smithsonian Astrophysical Observatory for and on behalf of NASA under contract NAS8-03060. %This work has made use of SAOImage DS9, developed by the Smithsonian Astrophysical Observatory \citep{joye06}, the {\tt XSPEC} spectral fitting package \citep{arnaud04}, the FUNTOOLS utilities package, the HEASARC FTOOLS package, and the CIAO (\cxo\ Interactive Analysis of Observations) package. The anonymous referee is thanked for providing comments which we are pleased to ignore.


\begin{thebibliography}{16}
%\bibitem[Arnaud(2004)]{arnaud04} Arnaud, K.\ 2004, Bulletin of the American Astronomical Society, 36, 934 
%\bibitem[Bernstein(2007)]{bern07} Bernstein, J.~P.\ 2007, Bulletin of the American Astronomical Society, 38, 997 
%\bibitem[Camilo et al.(2000)]{camilo00} Camilo, F., Kaspi, V.~M., Lyne, A.~G., Manchester, R.~N., Bell, J.~F., D'Amico, N., McKay, N.~P.~F., \& Crawford, F.\ 2000, \apj, 541, 367 
\bibitem[Borkowski et al.(2001)]{bork01} Borkowski, K.~J., Lyerly, W.~J., \& Reynolds, S.~P.\ 2001, \apj, 548, 820 
\bibitem[Dickey \& Lockman(1990)]{dl90} Dickey, J.~M., \& Lockman, F.~J.\ 1990, \araa, 28, 215 
\bibitem[F\"urst et al.(1985)]{fuerst85} F\"urst, E., Reich, W., Reich, P., Sofue, Y., \& Handa, T.\ 1985, \nat, 314, 720 
\bibitem[F\"urst et al.(1989)]{fuerst89} F\"urst, E., Hummel, E., Reich, W., Sofue, Y., Sieber, W., Reif, K., \& Dettmar, R.-J.\ 1989, \aap, 209, 361 
\bibitem[F\"urst et al.(1997)]{fuerst97} F\"urst, E., Reich, W., \& Aschenbach, B.\ 1997, \aap, 319, 655 
\bibitem[Gaensler \& Slane(2006)]{gaensler06} Gaensler, B.~M., \& Slane, P.~O.\ 2006, \araa, 44, 17 
\bibitem[Harrus et al.(2004)]{harrus04} Harrus, I.~M., Slane, P.~O., Hughes, J.~P., \& Plucinsky, P.~P.\ 2004, \apj, 603, 152 
%\bibitem[Joye(2006)]{joye06} Joye, W.~A.\ 2006, Astronomical Data Analysis Software and Systems XV, 351, 574 
\bibitem[Kondratiev et al.(2009)]{kondra09} Kondratiev, V.~I., McLaughlin, M.~A., Lorimer, D.~R., Burgay, M., Possenti, A., Turolla, R., Popov, S.~B., \& Zane, S.\ 2009, \apj, 702, 692 
%\bibitem[McLaughlin et al.(2003)]{mclaugh03} McLaughlin, M.~A., et al.\ 2003, \apjl, 591, L135 
\bibitem[Monet et al.(2003)]{monet03} Monet, D.~G., et al.\ 2003, \aj, 125, 984\bibitem[Nakanishi \& Sofue(2003)]{naka03} Nakanishi, H., \& Sofue, Y.\ 2003, \pasj, 55, 191 
\bibitem[Odegard(1986)]{ode86} Odegard, N.\ 1986, \aj, 92, 1372 
\bibitem[Patnaik et al.(1988)]{patna88} Patnaik, A.~R., Velusamy, T., \& Venugopal, V.~R.\ 1988, \nat, 332, 136 
\bibitem[Petre et al.(2002)]{petre02} Petre, R., Kuntz, K.~D., \& Shelton, R.~L.\ 2002, \apj, 579, 404 
\bibitem[Plucinsky et al.(2002)]{plu02} Plucinsky, P.~P., Dickel, J.~R., Slane, P.~O., Edgar, R.~J., Gaetz, T.~J., \& Smith, R.~K.\ 2002, APS Meeting Abstracts, 17037 
\bibitem[Possenti et al.(2002)]{possenti02} Possenti, A., Cerutti, R., Colpi, M., \& Mereghetti, S.\ 2002, \aap, 387, 993 
\bibitem[Posson-Brown et al.(2010)]{posson10} Posson-Brown, J., Grant, C., Allen, G., Plucinsky, P. P., \& Edgar, R. J.\ 2010, http://cxc.harvard.edu/twiki/bin/view/AcisCal/\\TCTI\_HowTo
%\bibitem[Rea et al.(2009)]{rea09} Rea, N., et al.\ 2009, \apjl, 703, L41 
%\bibitem[Rees \& Gunn(1974)]{rees74} Rees, M.~J., \& Gunn, J.~E.\ 1974, \mnras, 167, 1 
\bibitem[Reich et al.(1984)]{reich84} Reich, W., F\"urst, E., Haslam, C.~G.~T., Steffen, P., \& Reif, K.\ 1984, \aaps, 58, 197 
\bibitem[Safi-Harb(2004)]{safi04} Safi-Harb, S.\ 2004, Boletin de la Asociacion Argentina de Astronomia La Plata Argentina, 47, 277 
\bibitem[Seward \& Wang(1988)]{seward88} Seward, F.~D., \& Wang, Z.-R.\ 1988, \apj, 332, 199 
\bibitem[Smith et al.(2001)]{smith01} Smith, R.~K., Brickhouse, N.~S., Liedahl, D.~A., \& Raymond, J.~C.\ 2001, \apjl, 556, L91 
%\bibitem[Vink \& Bamba(2009)]{vinka09} Vink, J., \& Bamba, A.\ 2009, \apjl, 707, L148 
\bibitem[Toscano et al.(1999)]{toscano99} Toscano, M., Sandhu, J.~S., Bailes, M., Manchester, R.~N., Britton, M.~C., Kulkarni, S.~R., Anderson, S.~B., \& Stappers, B.~W.\ 1999, \mnras, 307, 925
\bibitem[Wilms et al.(2000)]{wilms00} Wilms, J., Allen, A., \& McCray, R.\ 2000, \apj, 542, 914
\end{thebibliography}
\end{document}